\documentclass[a4paper,11pt]{article}
\usepackage[T1]{fontenc} 
\usepackage[utf8]{inputenc} 
\usepackage{microtype} 
\usepackage{lmodern} 
\usepackage{authblk}
\usepackage{hyperref}
\usepackage{color}
\definecolor{dark-red}{rgb}{0.4,0.15,0.15}
\definecolor{dark-blue}{rgb}{0.15,0.15,0.6}
\definecolor{medium-blue}{rgb}{0,0,0.5}
\hypersetup{colorlinks, linkcolor={dark-red}, citecolor={dark-blue}, urlcolor={medium-blue}}
\usepackage[text={16.5cm,26cm},centering]{geometry}
\usepackage{authblk}
\let\OLDthebibliography\thebibliography
\renewcommand\thebibliography[1]{
  \OLDthebibliography{#1}
  \setlength{\parskip}{0pt}
  \setlength{\itemsep}{2pt plus 0.3ex}
}
\usepackage{amsmath,amsfonts,amssymb}
\usepackage{mleftright} \mleftright
\usepackage{tensor} 
\usepackage{mathrsfs} 
\newcommand*{\dd}{\mathop{}\!d}
\newcommand*{\email}[1]{\href{mailto:#1}{#1}}
\title{Aspects of AdS$_2$ holography with non-constant dilaton}
\author[1,2]{Daniel Grumiller\thanks{\email{grumil@hep.itp.tuwien.ac.at}}}
\author[1]{Jakob Salzer\thanks{\email{salzer@hep.itp.tuwien.ac.at}}}
\author[2]{Dmitri Vassilevich\thanks{\email{dvassil@gmail.com}}}
\affil[1]{\emph{Institute for Theoretical Physics, TU Wien}\\
\emph{Wiedner Hauptstrasse 8-10/136, A-1040 Vienna, Austria}}
\affil[2]{\emph{CMCC-Universidade Federal do ABC, Santo Andr\'e, S.P. Brazil}}
\date{}
\begin{document}
\maketitle
\vspace{-1.3 cm}
\begin{abstract}
In this proceedings contribution we summarize and discuss results of Refs.~\cite{Grumiller:2013swa, Grumiller:2015vaa} in the light of recent developments in $\textrm{AdS}_2$ holography \cite{Maldacena:2016upp, Jensen:2016pah, Engelsoy:2016xyb, Jevicki:2016bwu}. 
\end{abstract}
\vspace{-0.5cm}
\section{Introduction}
The application of new ideas and concepts to the most elementary reasonable set-up has proven a fruitful strategy in the past to unravel their inner workings. In this manner, the application of holography in the form of AdS/CFT to three-dimensional Einstein gravity brought a number of remarkable results, e.g.~the calculation of the Bekenstein--Hawking entropy using the Cardy formula \cite{Strominger:1997eq}.  However, at times the easiest set-up is not what one might naively expect. For instance, holography in two dimensions (2d) is more subtle than its higher-dimensional relative, partly because the global AdS$_2$ boundary is disconnected, partly because black hole horizons have no transversal space and partly because in most constructions the canonical boundary charges characterizing the physical states vanish identically.\par
Although Einstein gravity does not exist in 2d---the Einstein--Hilbert action yields just a topological invariant and no equations of motion---, the coupling of a scalar field, the dilaton $X$, to the metric yields a well-defined theory, 2d dilaton gravity, with rich classes of solutions. Its action is
\begin{equation}
  \label{eq:1}
  I=-\frac{k}{4\pi}\int_{\mathcal M}\dd^2x\sqrt{g}\left(X R-U(X)(\nabla X)^2-2V(X)\right)\,, 
\end{equation}
where  $U(X)$ and $V(X)$ are some arbitrary functions that define the particular model. A review of 2d dilaton gravity with an extensive list of references is presented in \cite{Grumiller:2002nm}.\par
In many cases the action \eqref{eq:1} admits $\textrm{AdS}_2$ solutions, which suggests to apply holographic methods to 2d dilaton gravity. 
Recently, $\textrm{AdS}_2$ holography has gained increased attention \cite{Maldacena:2016upp, Jensen:2016pah, Engelsoy:2016xyb,Jevicki:2016bwu} since some properties expected of a reasonable boundary theory are shared by the Sachdev-Ye-Kitaev model \cite{Sachdev:1992fk,Kitaev:15ur}. \par
In this proceedings contribution we discuss results previously obtained in \cite{Grumiller:2013swa, Grumiller:2015vaa} and highlight particular points in the light of the new developments mentioned in the previous paragraph. In section 2 we start with a brief summary of the gauge theory formulation of 2d dilaton gravity which has a status comparable to Chern--Simons theory in three dimensional gravity. In section 3 we discuss aspects of holography in constant and linear dilaton backgrounds. In section 4 we include remarks on recent developments. Section 5 contains a brief conclusion. \par
Our conventions follow \cite{Grumiller:2015vaa}: We work exclusively in Euclidean signature. Indices of elements of the frame bundle are raised and lowered using $\delta_{ab}=\textrm{diag}(1,1)_{ab}$, i.e. $e_a=\delta_{ab}e^b$. The volume form $\epsilon$ is defined as $\epsilon=\dd^2x\sqrt{g}=\frac{1}{2}\epsilon^{ab}e_a\wedge e_b=e_1\wedge e_0=*1$.
\section{Two-dimensional dilaton gravity as a non-linear gauge theory}
In the case of Einstein gravity in three dimensions, many calculations are considerably simplified in the Chern--Simons formulation. This is due to the trading of diffeomorphisms for pointwise gauge transformations. In a similar way, the realization of 2d dilaton gravity as a non-linear gauge theory \cite{Ikeda:1993fh} holds comparable simplification. In particular, 2d dilaton gravity arises as a specific Poisson sigma model (PSM) \cite{Schaller:1994es}. The transition is most lucid when the bulk action \eqref{eq:1} is rewritten in the first order formulation
\begin{equation}
  \label{eq:2}
  I=-\frac{k}{2\pi}\int_{\mathcal M}\left(X^aDe_a+X\dd \omega +\epsilon \mathcal{V}(X^c,X)\right)\,
\end{equation}
with the definition
\begin{equation}
  \label{eq:3}
  \mathcal{V}(X^c,X)=-\frac{1}{2}X^cX_cU(X)-V(X)\,.
\end{equation}
where $\omega$ is the dualized spin-connection and $e_a$ the zweibein. The covariant derivative $D$ is defined as
  $D e_a=d e_a+\epsilon\indices{_a^b}\omega\wedge e_b$.
The additional fields $X^a$ can be viewed as Lagrange multipliers enforcing the torsion constraint $D e_a=0$. 
\par
A PSM is a sigma model on a 2d manifold $\mathcal{M}$ with the target space $\Sigma$ being a Poisson manifold. Its action is given by
\begin{equation}
  \label{eq:5}
  I=-\frac{k}{2\pi}\int_{\mathcal M}\left(X^I\dd A_I+\frac{1}{2}P^{IJ}(X^K)A_I\wedge A_J\right)\,.
\end{equation}
Here $X^I$ denote coordinates on $\Sigma$, $A_I$ are 1-forms on $\mathcal{M}$ taking values in the cotangent space of $\Sigma$, and $P^{IJ}$ is the antisymmetric Poisson tensor on the target space that defines the Poisson bracket by
\begin{equation}
  \label{eq:20}
  \left\{X^I,X^J\right\}=P^{IJ}(X^K)\,.
\end{equation}
The Jacobi identity for the Poisson bracket translates to the the condition
  $(\partial_LP^{IJ})P^{LK}+(\partial_LP^{JK})P^{LI}+(\partial_LP^{KI})P^{LJ}=0$
on the Poisson tensor. \par
The action \eqref{eq:5} is invariant under the non-linear gauge transformations
\begin{subequations}
\begin{align}
  \label{eq:7}
  \delta_\lambda X^I&=P^{IJ}\lambda_J \\
\delta_\lambda A_I&=-\dd \lambda_I-\partial_IP^{JK}\lambda_KA_J\,.
\end{align}
\end{subequations}
up to a total derivative and yields the equations of motion
\begin{subequations}
\begin{align}
  \label{eq:8}
  \dd X^I+P^{IJ}A_J&=0\\
\label{eq:8b}
\dd A_I+\frac{1}{2}\partial_IP^{JK}A_J\wedge A_K&=0\,,
\end{align}
\end{subequations}
upon variation with respect to $X_I$ and $A_I$. \par
If one chooses a three-dimensional Poisson manifold parametrized by coordinates $(X,X^a)$ with Poisson tensor
\begin{equation}
  \label{eq:9}
  P^{ab}=\mathcal{V}\epsilon^{ab}\qquad P^{Xa}=X^b\epsilon\indices{_b^a}\,
\end{equation}
the two actions \eqref{eq:2} and \eqref{eq:5} coincide under the identification
  $A_X=\omega$, $A_a=e_a$.
The coupling of 2d dilaton gravity to (non-)abelian gauge fields is straightforward by adding further gauge connections and going to higher dimensional Poisson manifolds.  
\par
Since the Poisson tensor is antisymmetric and of odd dimension, it must be degenerate. Thus, there necessarily exists at least one Casimir function $\mathcal{C}$ with 
  $\left\{X^I,\mathcal{C}\right\}=P^{IJ}\frac{\partial\mathcal{C}}{\partial X^J}=0$.
The Poisson manifold decomposes into a union of symplectic leaves on each of which the Poisson tensor is non-degenerate and the Casimir functions are constant, cf. e.g. \cite{Marsden:1994ms}.\par
\par
The equations of motion \eqref{eq:8},\eqref{eq:8b} allow for two distinct sectors of solutions. Constant dilaton vacua (CDV) are defined by the condition that the $X^I$ are constant on-shell and the Poisson tensor vanishes
$X^K=\bar{X}^K$, $P^{IJ}(\bar{X}^K)=0$.
These solutions might or might not exist depending on the existence of roots of the potential $\mathcal{V}$. 
Generic solutions belong to the sector of linear dilaton vacua. These exist for any choice of the potential $\mathcal{V}$. 
In the following we study the linear dilaton sector for locally Euclidean $\mathrm{AdS}_2$ spacetimes. (The constant dilaton sector is holographically less interesting and is reviewed in \cite{Grumiller:2015vaa}.)

\section{Linear dilaton vacua in Euclidean locally $\textrm{AdS}_2$}
In the following we are going to restrict ourselves to the study of the Jackiw--Teitelboim model \cite{Jackiw:1984jt,Teitelboim:1984jt}. The defining functions are given by
\begin{equation}
  \label{eq:26}
  V=-X,\qquad U=0\,.
\end{equation}
This model has recently gained attention in the form of the Almheiri--Polchinski model with $V=\bar{X}-X$ \cite{Almheiri:2014cka}. It is clear that this model admits a CDV if $X$ is finetuned to $X=\bar{X}$. In the following we will stick to the potential \eqref{eq:26}. \par
Since we are interested in spacetimes that are locally Euclidean $\textrm{AdS}_2$, we impose the boundary conditions 
\begin{align}
  \label{eq:16}
  e_{1\rho}=1\qquad e_{1\varphi}=0\qquad e_{0\rho}=0\qquad e_{0\varphi}=\tfrac{1}{2}e^{\rho}-\tfrac{1}{2}M(\varphi)e^{-\rho}
\end{align}
on the zweibein with no further subleading components. The full geometry is specified by the `mass function' $M(\varphi)$. Compare this to the function $t(\varphi)$ appearing in \cite{Maldacena:2016upp}, that specifies the different cut-out geometries. \par
The connection 1-form is specified by requiring that the torsion constraint [the $0$ and $1$ components of \eqref{eq:8b}] holds exactly. This restricts $\omega$ to the form
\begin{equation}
  \label{eq:18}
  \omega_\rho=0\qquad \omega_\varphi=-\tfrac{1}{2}e^{-\rho}-\tfrac{1}{2}M(\varphi)e^{-\rho}\,.
\end{equation}
\par
The requirement $\partial_X\mathcal V=1$ for having a space of constant negative curvature $R=-2$ is obviously fulfilled by the potential \eqref{eq:26}. The boundary conditions on the dilaton field $X$ and the auxiliary fields $X^a$ are chosen as
\begin{align}
  \label{eq:15}
  X^0=X_R(\varphi) e^\rho-X_L(\varphi)e^{-\rho}\,,\qquad 
X^1=-2X_R'(\varphi)\,,\qquad
X=X_R(\varphi)e^{\rho}+X_L(\varphi)e^{-\rho}\,.
\end{align}
The Casimir function for this model is given by
  $\mathcal C=X_LX_R-\frac{1}{4}(X^1)^2$.
The boundary conditions \eqref{eq:16},\eqref{eq:18}, and \eqref{eq:15} are compatible with the equations of motion given that the following relation hold
\begin{equation}
  \label{eq:28}
  X_L=M X_R+2X_R''\qquad
  X_RM'+2X_R'M+2X_R''' = 0\,.
\end{equation}

A study of $\textrm{AdS}_2$ holography with these boundary conditions was presented in \cite{Grumiller:2013swa}. In our boundary conditions it is implicit that all the functions appearing are \emph{state-dependent}, in particular the dilaton field is allowed to vary to leading order. Therefore, the boundary condition preserving transformations are looser than might have been expected. Explicitly, they are given by
\begin{equation}
  \label{eq:22}
\lambda_0=\tfrac{1}{2}\lambda(\varphi)e^{\rho}-\left(\tfrac{1}{2}\lambda(\varphi)M(\varphi)+\lambda''(\varphi)\right)e^{-\rho}\qquad \lambda_1=-\lambda'(\varphi)\qquad
\lambda_X=\lambda_0-\lambda(\varphi)e^{\rho}\,.
\end{equation}
Using the relation $\lambda_{a}=\xi^\mu A_{\mu a}$ the corresponding diffeomorphisms read
\begin{equation}
  \label{eq:31}
\xi^{\rho}=-\lambda'(\phi)\qquad \xi^{\varphi}=\lambda-\frac{2 \lambda''}{e^{2\rho}-M(\phi)}
\end{equation}
Expanding $\lambda(\varphi)$ into Fourier modes and setting $T_n\equiv \xi[\lambda=e^{in\varphi}]$ we find that the asymptotic symmetry algebra corresponds to the Witt algebra
\begin{equation}
  \label{eq:32}
  \left[T_n,T_m\right]=(n-m)T_{n+m}\,.
\end{equation}
Letting the dilaton field vary to leading order is crucial for this finding.\footnote{ If the leading order of $X$ were held fixed, the symmetry would be broken to $U(1)$. This is the main difference between \cite{Maldacena:2016upp,Engelsoy:2016xyb} and our discussion.} 
Under these transformations the functions $X_R,M$ transform as
\begin{equation}
    \delta_\lambda X_R=X_R\lambda'-X_R'\lambda\qquad
\label{eq:33b}
    \delta_\lambda M=-M'\lambda-2M\lambda'-2\lambda'''\,.
\end{equation}
We see that the mass function $M(\varphi)$ transforms under an infinitesimal Schwarzian derivative. \par
The canonical boundary currents are non-zero
but in general non-integrable as noted in \cite{Grumiller:2013swa}. 
However, as shown in \cite{Grumiller:2015vaa} assuming that the zero mode of $X_R$ is a state independent quantity $\bar{X}$ is enough to integrate the zero-mode charge perturbatively, such that the normalization of the Virasoro algebra \eqref{eq:33b} can be fixed to
\begin{equation}
  \label{eq:29}
  \delta L=-L'\lambda -2L \lambda'-\frac{c}{12}\lambda'''
\end{equation}
with the central charge 
\begin{equation}
  \label{eq:34}
  c=24k\,\frac{\bar X}{2\pi}\,.
\end{equation}
Using the chiral Cardy formula thus yields the entropy
\begin{equation}
  \label{eq:35}
S_{\textrm{\tiny Cardy}} = \frac{\pi^2 c\, T}{3} = 2\pi\sqrt{\frac{c L}{6}} = 2k\bar X \sqrt{M} = 4\pi k\bar XT
\end{equation}
that coincides with the macroscopic calculation of entropy using Wald's method or the Euclidean path integral. 

\section{Remarks on recent developments}
The recent interest in $\textrm{AdS}_2$ holography was sparked by the finding that the boundary dynamics of $\textrm{AdS}_2$ is governed by an action that consists of a Schwarzian derivative \cite{Maldacena:2016upp,Jensen:2016pah,Engelsoy:2016xyb,Jevicki:2016bwu}, since a similar action appears in the analysis of the Sachdev--Ye--Kitaev model, which describes fermions with random interactions \cite{Sachdev:1992fk, Kitaev:15ur, Maldacena:2016hyu, Polchinski:2016xgd}. In the following we show how such a boundary action arises in the set-up presented
in this proceedings contribution. \par
According to the general case discussed in \cite{Bergamin:2007sm}, for a well-defined action principle the action \eqref{eq:5} has to be supplemented by the two boundary terms
\begin{equation}
  \label{eq:36}
  I_{\rm{\tiny bdy}}=\frac{k}{2\pi}\int_{\partial \mathcal{M}}\dd \varphi X
\left(\omega_\phi+e_\phi\right)\,,
\end{equation}
evaluated at some cut-off surface $\rho_c$. Imposing that the spacetime be locally $\textrm{AdS}_2$ and the connection torsion-free, the full action $I+I_{\rm{\tiny bdy}}$ is evaluated on the solutions \eqref{eq:16} and \eqref{eq:18}, and the bulk part vanishes
identically. The dilaton field is assumed to grow asymptotically as $X=X(\varphi)e^{\rho}+{\cal O}(1)$. When the cut-off surface is sent to infinity $\rho\rightarrow \rho_c$ the boundary part yields
\begin{equation}
  \label{eq:38}
  I=\frac{k}{2\pi}\int_{\partial \mathcal M} \dd \varphi
X(\varphi)M(\varphi)\,.
\end{equation}
This result \eqref{eq:38} is essentially the effective boundary action mentioned in the above works given the transformation property \eqref{eq:33b} of $M(\varphi)$ with an infinitesimal Schwarzian derivative. However, we stress that in our construction the dilaton $X$ is allowed to vary to leading order.

\section{Discussion and outlook}
\label{sec:discussion-outlook}
In this proceedings contribution we discussed 2d dilaton gravity in the first order formulation and reviewed the linear dilaton sector and its holographic discussion derived originally in \cite{Grumiller:2015vaa}. The asymptotic symmetry algebra is generically non-empty in the linear dilaton sector. In particular, in the case of the Jackiw--Teitelboim model, we find that the physical states form a representation of the Virasoro algebra given that the dilaton field is allowed to vary to highest order. \par
It would be interesting to further examine the similarities and differences between our set-up and the one discussed e.g. in \cite{Maldacena:2016upp,Engelsoy:2016xyb}. It appears to us that the main difference lies in allowing the leading order of the dilaton field to fluctuate in our construction. Furthermore, if the occurrence of a Virasoro algebra in \eqref{eq:29} is correct, one should reproduce a character of the Virasoro algebra in the one-loop contribution to the partition function. Work in this direction is currently in progress. 

\section*{Acknowledgments}

DG thanks Igor Klebanov for pointing out a talk based on Ref.~\cite{Maldacena:2016upp} during the MIAPP programme ``Higher-Spin Theory and Duality'' in May 2016 at the Munich Institute for Astro- and Particle Physics, Germany.

DG and JS were supported by the Austrian Science Fund (FWF), projects P 27182-N27 and P 28751-N27. DV was supported by CNPq, project 
306208/2013-0, and FAPESP, project 2012/00333-7.
DG and DV were additionally supported by the program Science without Borders, project CNPq-401180/2014-0.

\bibliography{AdS2Proceedings}

\providecommand{\href}[2]{#2}\begingroup\raggedright\begin{thebibliography}{10}

\bibitem{Grumiller:2013swa}
D.~Grumiller, M.~Leston, and D.~Vassilevich, ``{Anti-de Sitter holography for
  gravity and higher spin theories in two dimensions},'' {\em Phys. Rev.} {\bf
  D89} (2014), no.~4, 044001,
\href{http://www.arXiv.org/abs/1311.7413}{{\tt 1311.7413}}.

\bibitem{Grumiller:2015vaa}
D.~Grumiller, J.~Salzer, and D.~Vassilevich, ``{AdS$_{2}$ holography is
  (non-)trivial for (non-)constant dilaton},'' {\em JHEP} {\bf 12} (2015) 015,
\href{http://www.arXiv.org/abs/1509.08486}{{\tt 1509.08486}}.

\bibitem{Maldacena:2016upp}
J.~Maldacena, D.~Stanford, and Z.~Yang, ``{Conformal symmetry and its breaking
  in two dimensional Nearly Anti-de-Sitter space},''
\href{http://www.arXiv.org/abs/1606.01857}{{\tt 1606.01857}}.

\bibitem{Jensen:2016pah}
K.~Jensen, ``{Chaos and hydrodynamics near AdS$_2$},''
\href{http://www.arXiv.org/abs/1605.06098}{{\tt 1605.06098}}.

\bibitem{Engelsoy:2016xyb}
J.~Engels{\"o}y, T.~G. Mertens, and H.~Verlinde, ``{An investigation of
  AdS$_{2}$ backreaction and holography},'' {\em JHEP} {\bf 07} (2016) 139,
\href{http://www.arXiv.org/abs/1606.03438}{{\tt 1606.03438}}.

\bibitem{Jevicki:2016bwu}
A.~Jevicki, K.~Suzuki, and J.~Yoon, ``{Bi-Local Holography in the SYK Model},''
  {\em JHEP} {\bf 07} (2016) 007,
\href{http://www.arXiv.org/abs/1603.06246}{{\tt 1603.06246}}.

\bibitem{Strominger:1997eq}
A.~Strominger, ``{Black hole entropy from near horizon microstates},'' {\em
  JHEP} {\bf 9802} (1998) 009,
\href{http://www.arXiv.org/abs/hep-th/9712251}{{\tt hep-th/9712251}}.

\bibitem{Grumiller:2002nm}
D.~Grumiller, W.~Kummer, and D.~Vassilevich, ``{Dilaton gravity in
  two-dimensions},'' {\em Phys.Rept.} {\bf 369} (2002) 327--430,
\href{http://www.arXiv.org/abs/hep-th/0204253}{{\tt hep-th/0204253}}.

\bibitem{Sachdev:1992fk}
S.~Sachdev and J.-w. Ye, ``{Gapless spin fluid ground state in a random,
  quantum Heisenberg magnet},'' {\em Phys. Rev. Lett.} {\bf 70} (1993) 3339,
\href{http://www.arXiv.org/abs/cond-mat/9212030}{{\tt cond-mat/9212030}}.

\bibitem{Kitaev:15ur}
A.~Kitaev, ``{A simple model of quantum holography-2015. KITP strings
  seminar}.''

\bibitem{Ikeda:1993fh}
N.~Ikeda, ``{Two-dimensional gravity and nonlinear gauge theory},'' {\em Annals
  Phys.} {\bf 235} (1994) 435--464,
\href{http://www.arXiv.org/abs/hep-th/9312059}{{\tt hep-th/9312059}}.

\bibitem{Schaller:1994es}
P.~Schaller and T.~Strobl, ``{Poisson structure induced (topological) field
  theories},'' {\em Mod.Phys.Lett.} {\bf A9} (1994) 3129--3136,
\href{http://www.arXiv.org/abs/hep-th/9405110}{{\tt hep-th/9405110}}.

\bibitem{Marsden:1994ms}
J.~Marsden and T.~Ratiu, {\em Introduction to mechanics and symmetry}.
\newblock Springer-Verlag, 1994.

\bibitem{Jackiw:1984jt}
R.~Jackiw, ``{Liouville Field Theory: a Two-Dimensional Model for Gravity?},''
  in {\em {Quantum Theory of Gravity}}, {S. Christensen}, ed., pp.~403--420.
\newblock {Adam Hilger, Bristol}, {1984}.

\bibitem{Teitelboim:1984jt}
C.~Teitelboim, ``{The Hamiltonian Structure of Two-Dimensional Space-Time and
  its Relation with the Conformal Anomaly},'' in {\em {Quantum Theory of
  Gravity}}, {S. Christensen}, ed., pp.~327--344.
\newblock {Adam Hilger, Bristol}, {1984}.

\bibitem{Almheiri:2014cka}
A.~Almheiri and J.~Polchinski, ``{Models of AdS$_{2}$ backreaction and
  holography},'' {\em JHEP} {\bf 11} (2015) 014,
\href{http://www.arXiv.org/abs/1402.6334}{{\tt 1402.6334}}.

\bibitem{Maldacena:2016hyu}
J.~Maldacena and D.~Stanford, ``{Comments on the Sachdev-Ye-Kitaev model},''
\href{http://www.arXiv.org/abs/1604.07818}{{\tt 1604.07818}}.

\bibitem{Polchinski:2016xgd}
J.~Polchinski and V.~Rosenhaus, ``{The Spectrum in the Sachdev-Ye-Kitaev
  Model},'' {\em JHEP} {\bf 04} (2016) 001,
\href{http://www.arXiv.org/abs/1601.06768}{{\tt 1601.06768}}.

\bibitem{Bergamin:2007sm}
L.~Bergamin, D.~Grumiller, R.~McNees, and R.~Meyer, ``{Black Hole
  Thermodynamics and Hamilton-Jacobi Counterterm},'' {\em J. Phys.} {\bf A41}
  (2008) 164068,
\href{http://www.arXiv.org/abs/0710.4140}{{\tt 0710.4140}}.

\end{thebibliography}\endgroup
\bibliographystyle{fullsort}
\end{document}